\documentstyle[12pt]{article}

\def\journal{\topmargin .3in	\oddsidemargin .5in
	\headheight 0pt	\headsep 0pt
	\textwidth 5.625in 
	\textheight 8.25in 
	\marginparwidth 1.5in
	\parindent 2em
	\parskip .5ex plus .1ex		\jot = 1.5ex}
\journal

\catcode`\@=11
\def\marginnote#1{}
\newcount\hour
\newcount\minute
\newtoks\amorpm
\hour=\time\divide\hour by60
\minute=\time{\multiply\hour by60 \global\advance\minute by-\hour}
\edef\standardtime{{\ifnum\hour<12 \global\amorpm={am}%
	\else\global\amorpm={pm}\advance\hour by-12 \fi
	\ifnum\hour=0 \hour=12 \fi
	\number\hour:\ifnum\minute<10 0\fi\number\minute\the\amorpm}}
\edef\militarytime{\number\hour:\ifnum\minute<10 0\fi\number\minute}
\def\draftlabel#1{{\@bsphack\if@filesw {\let\thepage\relax
   \xdef\@gtempa{\write\@auxout{\string
      \newlabel{#1}{{\@currentlabel}{\thepage}}}}}\@gtempa
   \if@nobreak \ifvmode\nobreak\fi\fi\fi\@esphack}
	\gdef\@eqnlabel{#1}}
\def\@eqnlabel{}
\def\@vacuum{}
\def\draftmarginnote#1{\marginpar{\raggedright\scriptsize\tt#1}}
\def\draft{\oddsidemargin -.5truein
	\def\@oddfoot{\sl preliminary draft \hfil
	\rm\thepage\hfil\sl\today\quad\militarytime}
	\let\@evenfoot\@oddfoot	\overfullrule 3pt
	\let\label=\draftlabel
	\let\marginnote=\draftmarginnote
   \def\@eqnnum{(\theequation)\rlap{\kern\marginparsep\tt\@eqnlabel}%
\global\let\@eqnlabel\@vacuum}  }
\def\preprint{\twocolumn\sloppy\flushbottom\parindent 2em
	\leftmargini 2em\leftmarginv .5em\leftmarginvi .5em
	\oddsidemargin -.5in	\evensidemargin -.5in
	\columnsep .4in	\footheight 0pt
	\textwidth 10in	\topmargin  -.4in
	\headheight 12pt \topskip .4in
	\textheight 7.1in \footskip 0pt
	\def\@oddhead{\thepage\hfil\addtocounter{page}{1}\thepage}
	\let\@evenhead\@oddhead	\def\@oddfoot{}	\def\@evenfoot{} }
\def\numberbysection{\@addtoreset{equation}{section}
	\def\theequation{\thesection.\arabic{equation}}}
\def\underline#1{\relax\ifmmode\@@underline#1\else
	$\@@underline{\hbox{#1}}$\relax\fi}
\def\titlepage{\@restonecolfalse\if@twocolumn\@restonecoltrue\onecolumn
     \else \newpage \fi \thispagestyle{empty}\c@page\z@
	\def\thefootnote{\fnsymbol{footnote}} }
\def\endtitlepage{\if@restonecol\twocolumn \else \newpage \fi
	\def\thefootnote{\arabic{footnote}}
	\setcounter{footnote}{0}}  
\catcode`@=12
\relax
\def\figcap{\section*{Figure Captions\markboth
	{FIGURECAPTIONS}{FIGURECAPTIONS}}\list
	{Figure \arabic{enumi}:\hfill}{\settowidth\labelwidth{Figure 999:}
	\leftmargin\labelwidth
	\advance\leftmargin\labelsep\usecounter{enumi}}}
 \relax
\def\tablecap{\section*{Table Captions\markboth
	{TABLECAPTIONS}{TABLECAPTIONS}}\list
	{Table \arabic{enumi}:\hfill}{\settowidth\labelwidth{Table 999:}
	\leftmargin\labelwidth
	\advance\leftmargin\labelsep\usecounter{enumi}}}
 \relax
\def\reflist{\section*{References\markboth
	{REFLIST}{REFLIST}}\list
	{[\arabic{enumi}]\hfill}{\settowidth\labelwidth{[999]}
	\leftmargin\labelwidth
	\advance\leftmargin\labelsep\usecounter{enumi}}}
 \relax
\makeatletter
\newcounter{pubctr}
\def\publist{\@ifnextchar[{\@publist}{\@@publist}}
\def\@publist[#1]{\list
	{[\arabic{pubctr}]\hfill}{\settowidth\labelwidth{[999]}
	\leftmargin\labelwidth
	\advance\leftmargin\labelsep
	\@nmbrlisttrue\def\@listctr{pubctr}
	\setcounter{pubctr}{#1}\addtocounter{pubctr}{-1}}}
\def\@@publist{\list
	{[\arabic{pubctr}]\hfill}{\settowidth\labelwidth{[999]}
	\leftmargin\labelwidth
	\advance\leftmargin\labelsep
	\@nmbrlisttrue\def\@listctr{pubctr}}}
 \relax
\makeatother
\catcode`\@=11
\def\section{\@startsection {section}{1}{0pt}{-3.5ex plus -1ex minus
 -.2ex}{2.3ex plus .2ex}{\raggedright\large\bf}}
\catcode`\@=12
\newskip\humongous \humongous=0pt plus 1000pt minus 1000pt

\newif\ifdtup

\def\oldreffmt#1{\rlap{[#1]} \hbox to 2\parindent{}}

\def\figfmt#1{\rlap{Figure {#1}} \hbox to 1in{}}

\def\beq{\begin{equation}}
\def\eeq{\end{equation}}

\def\bea{\begin{eqnarray}}

\def\eea{\end{eqnarray}}
\catcode`\@=11
\def\eqnarray{\stepcounter{equation}\let\@currentlabel=\theequation
\global\@eqnswtrue
\global\@eqcnt\z@\tabskip\@centering\let\\=\@eqncr
\gdef\@@fix{}\def\eqno##1{\gdef\@@fix{##1}}%
$$\halign to \displaywidth\bgroup\@eqnsel\hskip\@centering
  $\displaystyle\tabskip\z@{##}$&\global\@eqcnt\@ne
  \hskip 2\arraycolsep \hfil${##}$\hfil
  &\global\@eqcnt\tw@ \hskip 2\arraycolsep $\displaystyle\tabskip\z@{##}$\hfil
   \tabskip\@centering&\llap{##}\tabskip\z@\cr}

\def\@@eqncr{\let\@tempa\relax
    \ifcase\@eqcnt \def\@tempa{& & &}\or \def\@tempa{& &}
      \else \def\@tempa{&}\fi
     \@tempa \if@eqnsw\@eqnnum\stepcounter{equation}\else\@@fix\gdef\@@fix{}\fi
     \global\@eqnswtrue\global\@eqcnt\z@\cr}

\catcode`\@=12
\font\tenbifull=cmmib10 
\font\tenbimed=cmmib10 scaled 800
\font\tenbismall=cmmib10 scaled 666
\relax

\def\thefootnote{\fnsymbol{footnote}}

\begin{document}
\begin{titlepage}
\begin{center}
June, 1992      \hfill       LBL-31066 \\
                        \hfill       UCB-PTH-91/34 \\
\vskip 0.1 in
{\large \bf Walking Technicolor and Electroweak Radiative Corrections}

\vskip .2 in
       {\bf  Raman Sundrum}
        \vskip 0.2 cm
       {\it Department of Physics\\
          University of California, Berkeley, CA 94720\\
          and\\
          Lawrence Berkeley Laboratory \\
          1 Cyclotron Road, Berkeley, CA 94720 \\   }
        \vskip 0.5 cm
  {
      {\bf  Stephen D.H. Hsu}\footnote{Junior Fellow, Harvard Society of
      Fellows} \\
           \vskip 0.2 cm
      }
        {\it Lyman Laboratory of Physics\\
             Harvard University\\
             Cambridge, MA 02138}
\end{center}
\vskip 0.05 in

\begin{abstract}
We examine the effect of walking technicolor dynamics on the
electroweak $S$ parameter and contrast it with the effect of QCD-like
technicolor dynamics. Our main tools are the
operator product expansion for the high-momentum behavior of the
electroweak gauge boson vacuum polarizations and the analyticity of these
polarizations which relate their low and high momentum behaviors. We
 show that whereas in large QCD-like technicolor models $S$ is large and
positive, in walking technicolor models a negative contribution is emphasized,
related to the large anomalous dimension of the technifermion condensate.
Thus in
walking technicolor $S$ is determined by a large cancellation of two competing
effects. This may result in  much smaller values of $S$ than  in
QCD-like technicolor, although
considerable uncertainties are involved. We conclude that it is impossible
to rule
out walking technicolor based on the present experimental limits on $S$ and the
present theoretical technology.
\end{abstract}
\end{titlepage}

\section{Introduction}
There has been a great deal of interest recently \cite{1,2} in possibly large
effects on precision electroweak observables due to isospin-preserving
 technicolor (TC) dynamics.
As noted by Peskin and Takeuchi \cite{2},
 the largest isospin-preserving effects from the
  electroweak symmetry breaking sector
 enter into physical observables through a single combination of  electroweak
vacuum polarizations which they call $S$.
(See refs. \cite{2} for the $S$ dependence of  precision observables.)
Although it has been appreciated for some time
that isospin violation in the electroweak Higgs sector
 is severely constrained by the experimental
value of the $\rho$ (or $T$) parameter, it is comparatively recently
that strong experimental limits have been placed on $S$. Experiments
 currently favor a
negative value of $S$.

Most discussions of the $S$ parameter (or closely related
parameters) within technicolor theories
 have focussed on what we will call `QCD-like'
technicolor (QTC). Such theories are taken to have dynamics very similar to
QCD,
supplemented by extended technicolor (ETC) four-fermion
interactions which  communicate the chiral symmetry
breaking of the strong
technicolor sector to ordinary fermions in order to give them masses
\cite{3}.
The  non-perturbative properties of such theories are estimated by the
appropriate
re-scalings of QCD properties. In particular, it is possible to compute $S$
in such
theories by organizing the calculation so that it depends on measured QCD
quantities \cite{2}. The result is that theories with large numbers of
technifermions give large, positive values of $S$, which are difficult to
reconcile with
experiment.
 The advantage of trying to use QCD-like dynamics in TC  theories
is that their properties are the
easiest to determine given our knowledge of QCD. The disadvantage is
that it is very difficult to construct models with ETC interactions
of sufficient strength to give realistic masses to the ordinary fermions,
while at the same time forbidding other ETC interactions which
mediate  unacceptably strong
 flavor changing neutral currents (FCNC's) \cite{4}.

Walking technicolor (WTC) \cite{5,5a} was proposed
 as a possible resolution of the problem of fermion masses and FCNC's. It was
shown
that a class of theories with slowly running (or  `walking')
technicolor couplings would produce enhanced technifermion condensates,
leading to  larger  masses for ordinary fermions, while
maintaining ETC mass scales sufficiently large to satisfactorily
suppress FCNC's. Whether or not fermion masses as large as that of the top
quark
can be accomodated within theories encorporating WTC is still an open question.
 WTC theories generally contain a
large number of technifermions, needed to slow the running of the technicolor
coupling, so  if calculations of $S$ performed
assuming QCD-like dynamics  were to apply to WTC,
unacceptably large values of $S$ would result.

In this paper we study the differences between WTC and QTC in their effects on
the $S$ parameter.
Our starting point will be the `dispersive approach' of Peskin and Takeuchi,
where $S$ is expressed as an integral over the TC spectrum in the isospin-one
vector and axial-vector channels, weighted towards the infrared.
Then, for QTC one can use experimental knowledge of the QCD spectrum to
estimate $S$.
Unfortunately, calculating the
 spectrum of a WTC theory is a strong interaction problem which we cannot
 solve.  Instead, we will
 apply the ACD (analytic continuation by duality) \cite{8} technique in order
to
      estimate $S$.

The central observation of
 the ACD method  relevant to our problem is that solving for the
 spectrum of the strongly interacting theory is solving a harder problem
 than necessary. After all, the Peskin-Takeuchi formula for  $S$ is a
 weighted mean of the TC spectrum and
 the details of the spectrum do not
 have to be known if one can get at $S$ directly.
To this end we apply the ACD
 method. First, $S$, a low-momentum quantity from the point of
 view of the TC theory, is expressed  in terms of the high-momentum (compared
 with the TC-scale)
 expansion of the electroweak vacuum polarizations. This expression follows
 is obtained by exploiting the   analyticity
 properties of the vacuum polarizations which relate high and low momenta.
The requisite high-momentum behavior is estimated by applying the operator
product
expansion (OPE) to products of electroweak currents.

 Our final result for the class of WTC and QTC models we consider is roughly
\begin{eqnarray}
S_{QTC} \sim (4.5 f_{TC}^2/m_C^2 - 0.015) N_D N_{TC} - 0.03 N_D, \nonumber \\
S_
{WTC} \sim (4.5 f_{TC}^2/m_C^2 - 0.07 c) N_D N_{TC} - 0.03 N_D.
\end{eqnarray}
Here, $N_D$ is the number of techni-doublets, $N_{TC}$ is the number of
technicolors,
$f_{TC}$ is the  decay constant of the eaten Goldstone particles
 of the theory ($246 {~\rm GeV}/\sqrt{N_D}$),
and $m_C$ is  the technifermion constituent mass.
$c$ is a positive constant which is of order one but whose exact value cannot
be determined. It is related to an OPE coefficient function.
The QTC result is robust because we can estimate $f_{TC}^2/m_C^2$
from QCD, and we find that the corresponding (positive) contribution to $S$
dominates. Therefore even if the resemblance to QCD is not perfect, we can
safely say that large QTC models of this sort will have large positive $S$,
$\sim 0.1 N_D N_{TC}$ (which agrees with \cite{2}).
  For WTC we cannot reach this conclusion. The  two largest
contributions, $4.5 f_{TC}^2/m_C^2 N_D N_{TC}$ and
$ - 0.07 c N_D N_{TC}$ have opposite sign and are of the same order, but they
are not precisely known.

Thus, $S_{WTC}$ is the result of a large, uncertain
cancellation and may be quite small.
 This makes it impossible at present to definitively rule out
 WTC theories. We believe that our estimate of $S_{WTC}$ is the best one can
do with the present understanding of WTC dynamics and without making
ad hoc assumptions about the details  of the WTC spectrum.
More than just the final answer though,
our approach makes manifest the link between the characteristic large anomalous
dimension of the technifermion bilinear in WTC and the potential for a
substantial reduction in $S$ over QTC.

The organization of the paper is as follows. In Section 2, we describe the
class of TC theories we are considering, define $S$, and derive the
Peskin-Takeuchi dispersive form for $S$ in our notation. In Section 3, the
ACD method is
applied to the problem of determining $S$ in TC, using the large-momentum
expansion of the electroweak polarizations as input. In
Section 4,  the general form of the large-momentum expansion is determined
using the  OPE. In Section 5,  we use the ACD method to calculate $S$ in the
simplest TC setting, minimal technicolor.
 In Section 6, the ACD method is applied to WTC and the result compared with
 QTC.  We present our
 conclusions in Section 7.

\section{The Peskin-Takeuchi formula}

In this paper, for the purpose of  discussing $S$,
we restrict our attention to  a class of simplified theories which we believe
retain the essential features of more realistic TC theories. We
consider $N_D$ identical doublets of massless technifermions, $\Psi_j = (U_j,
D_j)$, which
transform under $SU(2)_{EW} \times U(1)_Y$ and a vectorial technicolor gauge
symmetry, $SU(N_{TC})$, as follows:
\begin{eqnarray}
(U, D)_L &=& (2, 0, N_{TC}) \nonumber \\
U_R &=& (1, 1, N_{TC}) \nonumber \\
D_R &=& (1, -1, N_{TC}).
\end{eqnarray}
The strong TC force spontaneously breaks the technifermion chiral symmetries,
$SU(2N_D)_L \times SU(2N_D)_R \rightarrow SU(2N_D)_V$. This breaks
EW symmetry in
the correct way to give mass to the weak gauge bosons. Of the Goldstone bosons
produced, three are eaten by the weak gauge bosons while the remainder are
physical.  Their common Goldstone
decay constant must be $f_{TC} = 246~{\rm GeV}/\sqrt{N_{D}}$
 in order to ensure the correct masses for the electroweak gauge bosons.
Technifermion chiral symmetries are also explicitly broken by ETC
four-technifermion interactions. We consider a  simple form for the ETC
interactions which explicitly breaks the chiral symmetry
$SU(2N_D)_L \times SU(2N_D)_R \rightarrow SU(N_D)_V \times SU(2)_L \times
SU(2)_R$.
Ref. \cite{ss} gives the form of the ETC interactions which accomplish this,
although we will not need it explicitly in this paper. Here, $SU(N_D)_V$
symmetry reflects the fact that the simplified choice of ETC interactions does
not distinguish the $N_D$ technifermion doublets, $SU(2)_L$ is just the
gauged $SU(2)_{EW}$, and $U(1)_Y$ is a subgroup of $SU(2)_R$. In the absence of
electroweak interactions (but taking into account the ETC interactions)
there is a preserved symmetry group $SU(N_D)_V \times SU(2)_V$. $SU(2)_V$ is
the
custodial isospin symmetry  which protects the Peskin-Takeuchi $T$ parameter.
A realistic ETC theory would  violate this symmetry to some extent
 but since $T$ is not the focus of
this paper we have kept $SU(2)_V$ as a symmetry of the ETC and TC dynamics.
The explicit chiral symmetry breaking will give all the physical Goldstone
bosons a  common mass  (they are all in the same $SU(N_D)_V \times SU(2)_V$
multiplet),  making them pseudo-Goldstone bosons (PGB's).

In general, $S$ is defined in terms of the two-point function of the
technifermion
current coupled to the third component of weak isospin and the technifermion
current coupled to hypercharge. The electroweak interactions are neglected
within this correlator.
To be more precise, let $J_{\mu}^a$ and $J_{\mu}^Y$ denote the technifermion
$SU(2)_{EW}$ and $U(1)_Y$ currents respectively. Because they are conserved
their two-point function has the form,
\begin{equation}
i\int d^4x e^{i q.x} \langle 0 | T^*(J_{\mu}^3(x) J_{\nu}^Y(0)) | 0 \rangle
 = (q_{\mu}q_{\nu} - g_{\mu \nu} q^2) \Pi^{3Y}(s),
\end{equation}
where $s = q^2$.
$S$ is then defined as
\begin{equation}
S = -8 \pi (\Pi^{3Y}_{TC}(0) - \Pi^{3Y}_{MSM}(0)),
\end{equation}
where the second term is the analogous polarization for the minimal standard
model Higgs sector with a $1$ TeV Higgs mass. That is, $S$ measures deviations
from the minimal standard model predictions for precision observables.

To exploit the flavor symmetry of our class of TC models we
 define technifermion currents,
\begin{eqnarray}
V_{\mu}^{aj} &=& \overline{\Psi}^j \gamma_{\mu} \frac{\tau_a}{2} \Psi^j
\nonumber \\
A_{\mu}^{aj} &=& \overline{\Psi}^j \gamma_{\mu} \gamma_5 \frac{\tau_a}{2}
\Psi^j,
\end{eqnarray}
where $j$ is not summed over and labels one of the $N_D$ doublets, and $\tau_a$
are the Pauli matrices. The currents
defined above are triplets, $a = 1,2,3$, of custodial isospin.
 By Eq. (2),
\begin{eqnarray}
 J_{\mu}^a &=& \sum_{j} \frac{1}{2}(V_{\mu}^{aj} - A_{\mu}^{aj}), \nonumber \\
 J_{\mu}^Y &=& \sum_{j} (V_{\mu}^{3j} + A_{\mu}^{3j}).
\end{eqnarray}
By $SU(N_D)_V$ and parity invariance we then have,
\begin{equation}
 \langle 0 | T^*(J_{\mu}^3(x) J_{\nu}^Y(0)) | 0 \rangle =
\frac{1}{2} \sum_{j = 1}^{N_D} (\langle 0 | T^*(V_{\mu}^{3j}(x)
V_{\nu}^{3j}(0)) | 0 \rangle
- \langle 0 | T^*(A_{\mu}^{3j}(x) A_{\nu}^{3j}(0)) | 0 \rangle).
\end{equation}
These new two-point functions can also be expressed in the form
\begin{eqnarray}
i\int d^4x e^{i q.x} \langle 0 | T^*(V_{\mu}^{3j}(x) V_{\nu}^{3j}(0)) | 0
\rangle
 &=& (q_{\mu}q_{\nu} - g_{\mu \nu} q^2) \Pi^{V}(s), \nonumber \\
i\int d^4x e^{i q.x} \langle 0 | T^*(A_{\mu}^{3j}(x) A_{\nu}^{3j}(0)) | 0
\rangle
 &=& (q_{\mu}q_{\nu} - g_{\mu \nu} q^2) \Pi^{A}(s).
\end{eqnarray}
Again there is no summation over $j$ but nevertheless the right-hand sides are
independent of $j$ by the flavor symmetry. It is also useful to define
\begin{equation}
\Pi = \Pi^V - \Pi^A
\end{equation}
Then,
\begin{equation}
S = -4 \pi( N_D \Pi(0) - \Pi^{3Y}_{MSM}(0)).
\end{equation}
In the class of theories we will look at, working with $\Pi(t)$ will allow us
to look at contributions to $S$ {\em per doublet}, the only significant
$N_D$ dependence appearing explicitly as above.

Peskin and Takeuchi's `dispersive formula' for $S$ uses the
spectral representation for $\Pi(t)$ for complex
momentum-squared, $t$,
\begin{equation}
\Pi(t) = \frac{1}{\pi} \int_{0}^{\infty} ds \frac{{\rm Im}
\Pi (s)}{t - s + i \epsilon},
\end{equation}
This expresses the polarization $\Pi(t)$ in terms of Im$\Pi(s) = {\rm Im} \Pi_V
- {\rm Im} \Pi_A$, where ${\rm Im} \Pi_{V,A}$ are the technicolor spectra in
the isospin-one vector and axial channels.
The above integral is ultra-violet convergent as the vector and axial spectra
 cancel asymptotically. From Eq. (10) we obtain the
Peskin-Takeuchi formula,
\begin{equation}
S ~=~ 4  \int_{0}^{\infty} \frac{ds}{s}~ (N_D{\rm Im} \Pi_{TC}(s) -
 {\rm Im} \Pi^{3Y}_{MSM}(s)).
\end{equation}

Because the
electroweak interactions are turned off within the vacuum polarizations
appearing in the formula for $S$ the
true Goldstones are present and massless.
 The $\Pi$'s have been normalized so
that  the  Goldstone pole appears  in ${\rm Im} \Pi_A$ as
\begin{equation}
\frac{1}{\pi} {\rm Im} \Pi_A = f_{TC}^2 \delta(s) + ...~
\end{equation}
The integral over Im$\Pi(s)_{TC}$ in Eq. (12) diverges in the infrared because
of the true
Goldstone pole in the axial channel and the arbitrarily light
multi-Goldstone
states in both channels.  However these divergences are
eliminated by the
subtraction of the contributions of the analogous Goldstone states in the
standard model Higgs sector. We will say some more
about this subtraction in the next section. The massive PGB's
 do not contribute to the infrared divergence.

Now that we have specified what it is we wish to calculate, we proceed to
develop the means to do so.

\section{The ACD method}

In this section we  develop the formulas for calculating $S$ within
the class of theories we are considering. From the spectral representation, Eq.
(11), it is clear that $\Pi(t)$, even at large  $t$, depends on the
whole spectrum, Im$\Pi(s)$. In particular it depends on the low-energy
part of the spectrum. The ACD method attacks the inverse problem of calculating
quantities depending smoothly on the low-energy spectrum, like $S$, by
exploiting the behavior of $\Pi(t)$  for large $t$.
Below, we describe our adaptation of the ACD method for calculating $S$ in TC.

Our first task will be to immediately perform the MSM
 subtraction by subtracting the MSM spectrum from that of TC. The
 MSM spectrum is dominated by the Goldstone pole in the axial channel and the
 two-Goldstone states in the vector channel. The Higgs mass acts as a cutoff
 for these contributions. For all intents and purposes, this will coincide
with the TC spectrum below some sufficiently low scale $s_0$,
which we will fix later.
We will choose the Higgs mass-squared to be $s_0$. (In our later
 calculations this will turn out to be roughly $1~{\rm TeV}^2$, the value
 chosen by Peskin and Takeuchi in defining $S$. In any case $S$ is not very
 sensitive to this choice of the Higgs mass.)  Now the effect of the MSM
 subtraction can be  stated simply: the effective spectrum is that of the TC
theory above $s_0$ and  zero below. Thus $S$ can be written
\begin{equation}
S = 4 N_D \int^{\infty}_{s_0} \frac{ds}{s} {\rm Im} \Pi(s).
\end{equation}

Corresponding to the subtracted spectrum we define,
\begin{equation}
\overline{\Pi}(t) = \frac{1}{\pi} \int_{s_0}^{\infty} ds \frac{{\rm Im} \Pi
(s)}{t - s + i \epsilon},
\end{equation}
for which we wish to compute the large-$t$ behavior.
We will obtain the  large-$t$ behavior of $\Pi$ by using the OPE.
The general form will be
\begin{equation}
\Pi(t) \sim h_0 + h_1/t + h_2/t^2 + h_3/t^3 + ...~.
\end{equation}
where the $h_n$ depend weakly on
$t$. By noticing that
\begin{equation}
\overline{\Pi}(t) = \Pi(t) - \frac{1}{\pi} \int_{0}^{s_0} ds \frac{{\rm Im}
 \Pi (s)}{t - s + i \epsilon},
\end{equation}
the large-$t$ behavior for $\overline{\Pi}$ follows,
\begin{equation}
\overline{\Pi}(t) \sim h_0 + (h_1 - f_{TC}^2 + s_0/48 \pi^2)/t +
(h_2 + s_0^2/96 \pi^2)/t^2 + h_3/t^3 + ...~.
\end{equation}
That is, we have taken Eq. (16) and removed the contributions of the
Goldstone pole and the
two-Goldstone states below $s_0$ and expanded in $1/t$, where the couplings
to the currents are
determined in a chiral lagrangian. Other multi-Goldstone contributions make
negligble contributions.

Now we use the ACD trick \cite{8}.
 Let $R$ be a mass-squared large enough so that the above expansion makes
sense for $|t| \geq R$ and $t$ away from the physical axis.
Consider the two closed contours in Fig. 1.
By Eq. (15), $\overline{\Pi}$  is analytic in the regions
enclosed by both the contours. For any polynomial of finite degree $P(t) =
\sum_{n} a_n t^n$,
applying Cauchy's theorem to the integral of
$\overline{\Pi}(t) P(t)$ over $C = C_1 + C_2$  and
$\overline{\Pi}(t)/t$ over $D = D_1 - C_1$ we find, using Eq. (15),
\begin{eqnarray}
4 N_D \int_{s_0}^{R} ds {\rm Im}\Pi(s) P(s) &=&
 -2 i N_D \int_{C_1} dt \overline{\Pi}(t) P(t), \\
4 N_D \int_{R}^{\infty} ds \frac{{\rm Im}\Pi(s)}{s} &=&
2 i  N_D \int_{C_1} dt \frac{\overline{\Pi}(t)}{t},
\end{eqnarray}
The ACD technique consists of choosing $P(s)$ to approximate the
behavior of the weight function in question, in this case $1/s$,
 for real $s$  in the range $s_0$ to $R$. Then clearly by Eq. (14) the
 left-hand sides of
Eqs. (19,20)
 add up to $S$,
\begin{equation}
S \simeq 2 i N_D \int_{C_1} dt~ \overline{\Pi}(t) (1/t - P(t)).
\end{equation}
If we use the
expansion, Eq. (18), to approximate $\overline{\Pi}$ on $C_1$
and {\em if} we can neglect any $t$ dependence in the $h_n$, we can complete
$C_1$  to a
 circle in Eq. (21) and apply Cauchy's
formula, yielding,
\begin{eqnarray}
S &\simeq& 4 \pi N_D [h_0 - (h_1 - f_{TC}^2 + s_0/48 \pi^2)a_0 \nonumber \\
&-& (h_2 + s_0^2/96 \pi^2)a_1 - h_3 a_2 - ...].
\end{eqnarray}
(If the $t$ dependence of an $h_n$ is not negligibly weak  then one must keep
 it  and calculate
the right-hand side of Eq. (21)  without the aid of
Cauchy's formula.) Note  that  the integrand in Eq. (21) contains the factor
$(1/t - P(t))$. Because $P(t)$ is chosen to
make this factor approximate zero {\em on the physical axis}, the contour
integral representation for $S$ is dominated by $t$ far away from the axis.
This is where the OPE is expected to be most trustworthy. We will illustrate
this point later.

\section{The operator product expansion}

We now turn to the large-$t$ expansion for $\Pi(t)$. The important tool here is
the OPE for the product of currents in the vacuum polarization.
 The  product of currents transforms in the (1,3,3) representation
of $SU(N_D)_V \times SU(2)_L \times SU(2)_R$, a symmetry of the dynamics,
so all local
operators appearing in its OPE  must
carry the same representation (whether or not the vacuum is symmetric)
 \cite{11}. This is a strong constraint on the form of the OPE.

To begin, we neglect the ETC interactions in the theory and concentrate
on the TC
dynamics.
Then, for large spacelike momentum  $q^2 ~=~ -Q^2 ~<~ 0$, $\Pi(Q^2)$ has
the form
\begin{eqnarray}
\Pi(Q^2) &\sim& C_6(Q/\mu, \alpha(\mu))
\frac{\langle\overline{\psi} \psi\rangle^2_{\mu}}{Q^6} \nonumber \\
 &+& C_8(Q/\mu, \alpha(\mu))
\frac{\langle\overline{\psi} F_{\alpha \beta} \sigma^{\alpha
\beta}\psi\rangle_{\mu}
\langle\overline{\psi} \psi\rangle_{\mu}}{Q^8} \nonumber \\
&+& C_{10}(Q/\mu, \alpha(\mu))
\frac{\langle\overline{\psi} F_{\alpha \beta}
\sigma^{\alpha \beta}\psi\rangle_{\mu}^2}{Q^{10}} \nonumber \\
 &+& C_{10}'(Q/\mu, \alpha(\mu))
\frac{\langle\overline{\psi} F_{\alpha \beta} F^{\alpha \beta}\psi\rangle_{\mu}
\langle\overline{\psi} \psi\rangle_{\mu}}{Q^{10}} + ...,
\end{eqnarray}
where $F_{\alpha \beta}$ is the technigluon field strength, $\psi$ represents
any fermion flavor (which flavor is unimportant by the flavor symmetry)
  and $\mu$ is the
arbitrary renormalization scale.  The vacuum expectations of the operators
appearing in the OPE have been factorized in a vacuum dominance approximation
into products of technifermion bilinears.
 This  becomes exact in leading order in a
large-$N_{TC}$ approximation. The standard large-$N_{TC}$ approximation
suppresses technifermion loops which  are crucial for the walking of the gauge
coupling in WTC theories. In appendix 1 we describe the modification of
large-$N_{TC}$ counting neccesary to accomodate walking dynamics. Vacuum
dominance continues to hold in the modified scheme.
This is the most general expansion subject
to the
$SU(N_D)_V \times SU(2)_L \times SU(2)_R$ symmetry.

It is convenient to express the condensates
involving the technigluon fields as a product of a `default' estimate,
multiplied by a possible correction factor. For QCD-like theories at least it
is natural to guess that
\begin{eqnarray}
\langle\overline{\psi} F_{\alpha \beta} \sigma^{\alpha
\beta}\psi\rangle_{\mu} &\sim&
\sqrt{N_{TC}}\Lambda_{TC}^2 \langle\overline{\psi} \psi\rangle_{\mu}, \nonumber
\\
\langle\overline{\psi} F_{\alpha \beta} F^{\alpha \beta}\psi\rangle_{\mu}
&\sim&
N_{TC}^2  \Lambda_{TC}^4 \langle\overline{\psi} \psi\rangle_{\mu}.
\end{eqnarray}
Using this and choosing  $\mu = Q$, leads to an  expansion of the form
\begin{eqnarray}
\Pi(Q^2) &\sim& C_6(\alpha(Q))~
\frac{\langle\overline{\psi} \psi\rangle^2_{Q}}{Q^6} \nonumber \\
 &+& a(Q) C_8(\alpha(Q))
\frac{\sqrt{N_{TC}}m_C^2 \langle\overline{\psi} \psi\rangle^2_{Q}}{Q^8}
\nonumber \\
&+& b(Q) C_{10}(\alpha(Q))~
\frac{N_{TC} m_C^4 \langle\overline{\psi} \psi\rangle^2_{Q}}{Q^{10}} \nonumber
\\
&+& b'(Q) C'_{10}(\alpha(Q))~
\frac{N_{TC}^2 m_C^4 \langle\overline{\psi} \psi\rangle^2_{Q}}{Q^{10}}
\end{eqnarray}
The coefficients $a(Q), b(Q), b'(Q)$ are the correction factors referred to
above.

\section{Minimal TC}

In order to  illustrate and verify our procedure in a simple setting we first
compute $S$ for minimal technicolor (MTC). These are the QTC models with $N_D =
1$, so there are no PGB's.

Calculating the coefficient
functions perturbatively (and in large-$N_{TC}$ approximation) leads to the
expansion
\begin{eqnarray}
\Pi(Q^2) &\sim& - 4 \pi~ \alpha(Q)~
\frac{\langle\overline{\psi} \psi\rangle^2_{Q}}{Q^6} \nonumber \\
 &+& a(Q) (4 \pi~ \alpha(Q))^{3/2}
\frac{\sqrt{N_{TC}}m_C^2 \langle\overline{\psi} \psi\rangle^2_{Q}}{Q^8}
\nonumber \\
&+& b(Q)  (4 \pi~ \alpha(Q))^{2}
\frac{N_{TC} m_C^4 \langle\overline{\psi} \psi\rangle^2_{Q}}{Q^{10}}  + ...
\end{eqnarray}
In MTC we expect $a(Q), b(Q) \sim O(1)$, which agrees with phenomenological QCD
estimates \cite{iof}.
The $Q$ dependence of the technifermion condensate is easily determined
by the perturbative anomalous dimension,
\begin{equation}
\langle\overline{\psi} \psi\rangle_{Q} \sim
\langle\overline{\psi} \psi\rangle_{2 m_C}
(\frac{\alpha(2m_C)}{\alpha(Q)})^{9/22}.
\end{equation}
{}From QCD we estimate,
\begin{equation}
\langle\overline{\psi} \psi\rangle_{2 m_C} \sim - 0.12 N_{TC} m_C^3.
\end{equation}
(For  $N_C = 3$, $m_C = 350 MeV$, this gives
$\langle\overline{\psi} \psi\rangle_{2 m_C} \sim -(250 {\rm MeV})^3$.)
Using this and continuing to complex $t$,
\begin{equation}
\Pi(t) \sim 0.4 N_{TC} m_C^6/t^3 + 2 a(t) N_{TC} m_C^8/t^4 +
15 b(t) N_{TC} m_C^{10}/t^5 + ...
\end{equation}
The non-canonical $t$ dependence in the dominant term is extremely weak and we
have therefore neglected it.  We took $\alpha(2 m_C) \sim 2.5/N_{TC}$.

We must now choose $s_0$ and $R$. Judging from QCD, the non-negligible tail of
the techni-$\rho$ persists down to $\sim \frac{1}{2} m_{\rho}^2 \sim 2.5
m_C^2$. Below this however the MTC spectrum is given by the Goldstone states.
Therefore we choose $s_0 = 2.5 m_C^2$. We will trust the expansion to make
sense above $R = 25 m_C^2$ (in QCD this would be $3 \rm ~GeV^2$). Between $s_0$
and $R$ we have the approximation $1/t \sim P(t)$, where
\begin{eqnarray}
P(t) &=& 0.65/m_C^2 - 0.14 t/m_C^4 + 0.013 t^2/m_C^6 \nonumber \\
&-& 0.0006 t^3/m_C^8 + 8 \times 10^{-6} t^4/m_C^{10}.
\end{eqnarray}
A comparison is made in Fig. 2. A higher degree polynomial will only
 emphasize the unknown higher dimension condensate contributions in Eq. (22)
 and not improve our final accuracy.

The ACD formulas of Section 3 now yield
\begin{equation}
S = (8.2 f_{TC}^2/m_C^2 - 0.065 \pm 0.015) N_{TC} - 0.03,
\end{equation}
where the $\pm 0.015 N_{TC}$ is due to the uncertainty incorporated in $a, b$.
The last ingredient is the estimate from QCD,
\begin{equation}
f_{TC}^2 = 0 .024 N_{TC} m_C^2.
\end{equation}
 Putting this in,
\begin{equation}
S_{MTC} \sim (0.13  \pm 0.015) N_{TC} - 0.03.
\end{equation}
 The other uncertainty in this computation is the fit error between $P(t)$ and
 $1/t$ which we will not discuss because MTC
is not the main focus of this paper. It is small and the reasoning is very
similar to that given for the WTC case in appendix 2. The term subleading in
$1/N_{TC}$ is due
to MSM-subtracted two-Goldstone states. Although formally subleading, it
is not quite
negligible for moderate $N_{TC}$. Our result can be compared with the
estimates of Peskin and Takeuchi \cite{2}
\begin{equation}
S_{MTC} \sim 0.22, 0.32, 0.45,
\end{equation}
for $N_{TC} = 2, 3, 4$ respectively.

An interesting feature of  our calculation is that the dominant
contribution to  $S_{MTC}$ comes from the $f_{TC}^2/m_C^2$ term of the
$\overline{\Pi}(t)$ expansion. The appearance of this term
is easy to understand. The fact that chiral symmetry forbids the appearance of
a $1/t$ term  in the large-$t$ expansion of $\Pi(t)$ can be used to prove
the first Weinberg sum rule \cite{11,13}. This states that the total
weight in the vector channel is exactly the same as the total weight in
the axial channel. However in calculating $S$ we remove the Goldstone pole as
part of the MSM subtraction. This means that the effective weight in the
 axial channel is reduced by $f_{TC}^2$,  favoring positive $S$. The
 information we obtain from the non-perturbative condensates concerns
 the distributional aspects of the spectral weights in the two channels.

\section{Walking technicolor}

We now examine $S$ in WTC. We first need to discuss the following ingredients:
(i) the OPE coefficient functions, (ii) the condensates, (iii)
 the effects of the ETC interactions, (iv) the value of $f_{TC}^2/m_C^2$.

(i) An important observation about the large-momentum
expansion, Eq. (25), is that the coefficient functions $C_i$ all depend on $Q$
 only through $\alpha(Q)$. Therefore this dependence is very weak in WTC since
 $\alpha$ runs very slowly, so
 we will treat the $C_i$ as constants from here on.
In a WTC theory with anomalous dimension $\sim 1$ for $\overline{\psi} \psi$,
 ladder diagram analysis suggests that
\begin{equation}
\alpha(Q)_{WTC} \sim 2 \pi/(3 N_{TC}),
\end{equation}
 in the
walking region above $2 m_C$ \cite{5a}. Thus unlike MTC and QCD, the coupling
does not
rapidly become perturbatively small as $Q$ increases. This means the constants
$C_i$ cannot be very accurately determined by perturbative means
 because higher order corrections are not suppressed.
But we can at least say that the exact $C_i$ are  naturally  of the same
order as the leading perturbative estimates for them.
 For  $C_6$ we can also determine the sign, which is important. Because
WTC is a vector-like gauge theory, $\Pi(Q^2)$ is negative for all $Q^2 > 0$
\cite{witten}. This can be seen in Eq. (11) for spacelike momentum, $t = -
Q^2$. We will be considering $Q^2$ large enough so that  the
large-momentum expansion should be reliable and the dominant term appearing
in it
 determines the sign of $\Pi(Q^2)$. Thus $C_6 < 0$.
We will show that this corresponds to
a substantial negative contribution to $S_{WTC}$.

(ii) The key dynamical input from WTC is the renormalization-scale dependence
 of the technifermion
condensate. Let us consider the extreme
situation in which the anomalous
dimension of $\overline{\psi} \psi$ is one. The picture of the technifermion
self-energy $\Sigma(p)$ is then  \cite{5a},
\begin{eqnarray}
\Sigma(p) &\sim& m_C,~~ |p| < 2 m_C, \nonumber \\
&\sim& m_C^2/p,~~ |p| > 2 m_C ~{\rm and~ in~ the~ walking~ region}, \nonumber
\\
&\sim& m_C^3/p^2,~~ |p| ~{\rm so~ large~ that}~ \alpha_{TC}(p) ~{\rm is~
perturbatively~ small}.
\end{eqnarray}
Therfore for moderate  $\mu > 2 m_C$,
\begin{eqnarray}
\langle\overline{\psi} \psi\rangle_{\mu} &=& - \frac{N_{TC}}{2 \pi^2}
\int^{\mu}_{0} dp ~p^3~
\frac{\Sigma(p)}{p^2 +
 \Sigma(p)^2} \nonumber \\
     &\sim& \langle\overline{\psi} \psi\rangle_{2 m_C}
 - \frac{N_{TC}}{2 \pi^2} (\mu - 2m_C) m_C^2 \nonumber \\
&\sim& \langle\overline{\psi} \psi\rangle_{2 m_C} (\frac{\mu}{2 m_C}).
\end{eqnarray}
  We  borrowed the QCD estimate,
Eq. (28) because below $2 m_C$ the technifermions
decouple from the theory and the
dynamics should be similar to QCD. We see
that there is a linear enhancement of the condensate with renormalization
scale, $\mu$. It is this enhancement which is responsible for raising
ordinary fermion masses in a WTC theory.
 Less extreme forms of enhancement are
also possible in WTC theories. We comment on these possibilities in our
conclusions.

We must also decide whether to trust the estimates, Eq. (24), for the
 condensates involving the technigluon fields. The hint of
the large walking anomalous dimension of the  pure technifermion
bilinear can be found in the  large perturbative anomalous behavior, Eq. (27),
 present in  a running theory. But perturbatively,
 the mixed condensate has a small  anomaluous behavior \cite{nar}, suggesting
 that it is probably not enhanced to the same degree in WTC as the
 technifermion bilinear.
  Therefore Eq. (24) seems to be overestimating in ascribing the same
  enhancement to the mixed condensates as for the technifermion condensate. To
  be conservative we will take it as an  upper bound on the mixed
  condensates in order to make our
exploratory calculation of $S_{WTC}$.

(iii) As yet we have not included the effects of the ETC four-technifermion
operators which
give masses to the PGB's. Their contributions in the OPE can be calculated to
first order in the ETC interaction to give
\begin{equation}
\delta_{ETC} \Pi(t) \sim - m_{PGB}^2 f_{TC}^2/t^2,
\end{equation}
where we have expressed the strength of the ETC operator in terms of
 $m_{PGB}^2$, using
chiral perturbation theory \cite{5a,ss}. It is a noteoworthy feature of WTC
theories
that PGB masses can be quite large \cite{5a}. In this paper we will consider
\begin{equation}
m_{PGB} = m_C.
\end{equation}
Other possibilities are briefly discussed in our conclusions.

(iv) As in the MTC case, $S$ in WTC will be quite sensitive to
$f_{TC}^2/m_C^2$ for the reason that this determines how much less spectral
weight is available to contribute from the axial channel than from the vector
channel. Unfortunately, all that is really known is that it is of the same
order as in (large-$N_C$ rescaled) QCD \cite{5,5a},
\begin{eqnarray}
(f_{TC}^2/m_C^2)_{WTC} =  0.024 g N_{TC},
\end{eqnarray}
where $g \sim O(1)$, and we have used the estimate from QCD, Eq. (32).

The final version of the large-$t$
expansion  based on all the above general considerations reads
(continuing to complex $t$)
\begin{equation}
\Pi_{WTC}(t) \sim - f_{TC}^2 m_C^2 /t^2
- 0.1 c N_{TC} m_C^4/t^2 + 0.5 d(t) N_{TC} m_C^6/t^3
+ 5 e(t) N_{TC} m_C^8/t^4 + ...
\end{equation}
The first term is due to the ETC perturbation. $c$ is a positive coefficient
of order one, parameterizing our uncertainty in $C_6$. $d(t), e(t)$ are
correction factors for the estimates of the technigluon-technifermion
condensates
and the higher OPE coefficients. As explained in (ii),  it is likely
that Eq. (24) is a good bound on these condensates,
which corresponds to
\begin{equation}
|d(t)|, |e(t)| \leq 1.
\end{equation}
 We can now proceed to the ACD calculation.

The choice of $s_0$ is clear. Since the lightest states, other than true
Goldstone bosons, coupling to the electroweak currents consist of two PGB's,
we can take $s_0 = (2m_{PGB})^2 = (2m_C)^2$. We will trust our expansion at
$R = 25 m_C^2$.
In the range $s_0$ to $R$ we will use the approximation $1/t \sim  P(t)$, where
\begin{equation}
P(t) = 0.41/m_C^2 - 0.054 t/m_C^4 + 0.0028 t^2/m_C^6  - 5.1
\times 10^{-5} t^3/m_C^8.
\end{equation}
This is shown in Fig. 3. We emphasize again that the factor $1/t - P(t)$ which
appears in the integrand of Eq. (21) strongly
weights the integral near the deep Euclidean region
where the OPE is expected to be most trustworthy, as shown in Fig. 4.
There will be an error in calculating $S$ by the ACD technique due to the fact
that $P(t)$ is not exactly $1/t$ in the range $s_0$ to $R$. We estimate a bound
on this error in appendix 2 and find it to be quite small. The
significant uncertainties in $S$ are not inherent to the  ACD technique but
are due to the uncertainties in the input large-$t$ expansion in WTC.
If the first few terms in
this expansion were known precisely the ACD technique would give quite a
reliable estimate for $S$.

$S$ can now be computed. The result is
\begin{equation}
S_{WTC} = 4.5 N_D f_{TC}^2/m_C^2  - (0.07 c \pm 0.02)N_D N_{TC} - 0.03 N_D.
\end{equation}
The $\pm 0.02 N_D N_{TC}$ is due to the  uncertainty in $d, e$, which can
clearly be absorbed into  $c$. Using Eq. (40)
for $f_{TC}^2/m_C^2$, leads to our central result,
\begin{equation}
S_{WTC} = (0.11 g  - 0.07 c) N_D N_{TC} - 0.03 N_D,
\end{equation}
where $g$ and $c$ are order-one positive coefficients.

 To get a feel for the difference that
walking dynamics makes, let us contrast the above result with that for QTC
with PGB's of the same mass. The OPE is very similar to that of MTC except the
ETC perturbation must be incorporated,
\begin{equation}
\Pi_{QTC}(t) \sim - f_{TC}^2 m_C^2 /t^2 + 0.4 N_{TC} m_C^6/t^3 +
2 a(t) N_{TC} m_C^8/t^4 + 15 b(t) N_{TC} m_C^{10}/t^5 + ...
\end{equation}
Using this and the same $s_0$ and $R$ as before,
 we obtain
\begin{equation}
S_{QTC} = 4.5 N_D f_{TC}^2/m_C^2 - (0.014 \pm 0.001)N_D N_{TC} - 0.03 N_D.
\end{equation}
The $\pm 0.001$ is due to the uncertainty in $a, b$. Using the QCD estimate Eq.
(32)
for $f_{TC}^2/m_C^2$,
\begin{equation}
S_{QTC} \sim 0.095 N_D N_{TC} - 0.03 N_D.
\end{equation}

We see that in QTC, the overwhelmingly dominant contribution to $S$ is made
by the
$f_{TC}^2/m_C^2$ term. The leading condensate in the large-$t$ expansion of
$\Pi(t)$ does very little.
However, in WTC the leading term
in the expansion (without the ETC perturbation) has been promoted by the
large condensate anomalous
dimension from a canonically $1/t^3$ term to a $1/t^2$ term. The vanishing of
the $1/t^2$ term {\em asymptotically}
leads to the second Weinberg sum rule for the
spectrum \cite{}.
In WTC this sum rule remains true, but the persistence of the $1/t^2$ behavior
in the walking regime shows that saturating the second Weinberg sum rule with
the low-energy WTC spectrum is invalid.
In the ACD formulas of Eqs. (21,22) this promotion is very important, and one
can see
that it corresponds to a hefty negative contribution to $S_{WTC}$,
$- 0.07 c N_D N_{TC}$. This is where the sign of $C_6$ is important, which we
argued on general grounds had to be negative. If it had been positive the
contribution to $S$ would have been positive.

 So, large QTC models
 certainly have large positive $S$, as reported in the literature \cite{1,2},
{\em
 regardless of any small deviations from an exact QCD resemblance}. This is to
 be contrasted with WTC which is determined by a  cancellation of two large and
uncertain contributions, as we see in  Eq. (45). All we really know
about $g$ and $a$ is that they are
order one and positive.
Evidently, WTC as described
here may or may not be consistent with the current experimental
limits on $S$, depending
sensitively on $f_{TC}^2/m_C^2$ and $C_6$.

\section{Conclusions}

We used the analyticity of the electroweak polarizations
to relate their high and low momentum behavior,  and the OPE to capture the
essential
dynamics at high momentum. This led to a formula for $S$
which made transparent the relationship between $S$ and the large anomalous
dimensions of the technifermion condensate in walking technicolor.
Whereas in large QCD-like technicolor models $S$ is
necessarily large and positive, in walking technicolor there is an
enhancement of a negative
contribution so that $S$ is determined by a large and uncertain cancellation
of two competing effects. Because of this  it is impossible to say whether or
not walking technicolor theories are
 ruled out by the present experimental limits on $S$.

We have tried to present a formulation of the theoretical determination of
$S$, and in particular  the uncertainties involved, without resort to
ad hoc assumptions about the details of the walking technicolor spectrum.
 We believe
that with the present theoretical understanding of walking technicolor, $S$
cannot be determined
to greater precision than we have given without making such unjustified
assumptions.

Finally we consider how our results would change if (a) some of the PGB's are
lighter than we took them to be, or (b) if the walking technicolor model had
anomalous dimension
smaller than one for the technifermion bilinear. In both cases $S$ will
increase.

(a) The contributions to $S$ of two-PGB states lighter than $s_0 = (2 m_C)^2$
  will be positive as they occur in the vector channel.
These contributions can be
estimated using a  chiral lagrangian with a cutoff of $2 m_C$ (see
references \cite{1}).

(b) As the anomalous dimension of the fermion bilinear
varies between small values and one, the dynamics interpolates between
essentially  QCD-like technicolor and the extreme form of WTC we considered
here, so $S$ should vary accordingly. Thus
walking technicolor theories with anomalous dimension less than one would
 presumably have
larger values of $S$.

If these features are present in realistic walking technicolor theories
then $S_{WTC}$
 is most likely at least positive. If
experiments continue to converge on a large negative value of $S$, such
theories
are not likely to be realized in nature.

Recently Appelquist and Triantaphyllou have also estimated $S$ in WTC using a
`non-local chiral lagrangian' to capture the essential dynamics \cite{triant}.

\section*{Acknowledgements}

We thank R.S. Chivukula, L. Hall, M.A. Luty, S. Narison, M. Peskin,
U. Sarid, M. Soldate, M. Suzuki,
 T. Takeuchi, J. Terning and G. Triantaphyllou for
useful conversations. R.S. was supported in part by the Director, Office
of Energy Research, Office of High Energy and Nuclear Physics, Division of High
Energy Physics of the U.S. Department of Energy under contract
DE-AC03-76SF00098 and in part by the National Science Foundation under grant
PHY90-21139. S.D.H was supported by the National Science Foundation under grant
NSF-PHY-87-14654 and the state of Texas under grant PNRLC-RGFY106.

\section*{Appendix 1}

In order to describe walking dynamics in a large-$N_{TC}$ framework the number
of technifermion doublets is taken to increase proportionately to $N_{TC}$.
This means that  technifermion loops which are crucial for making the TC
gauge coupling `walk' are not forbidden in diagrams at the leading order of
$1/N_{TC}$. This is the only modification of standard large-$N_{TC}$
diagrammatics needed in WTC. The virtue of this approximation is that the
four-technifermion condensates appearing in the OPE expansion for $\Pi$
factorize in the leading order of $1/N_{TC}$ into two fermion bilinear
condensates. The crucial observation is  that
 all non-factorizeable contributions
are obtained from a factorizeable contribution by connecting  the two factors
with at least one technigluon. This makes them $O(1/N_{TC}^2$ smaller than the
factorized contributions.
 In this large-$N_{TC}$ approximation we have written all
the operators which can occur in the OPE, using the equations of motion acting
on the vacuum state to eliminate operators whenever possible.

A discussion of WTC dynamics in this large-$N_{TC}$ approximation is given in
ref. \cite{me}.

\section*{Appendix 2}

Here, we address the question of the fit error. The
polynomial $P(t)$ is not exactly the same as $1/t$ between $s_0$ and
$R$. Therefore there is a fit error in computing $S$ given by
\begin{equation}
 \delta S = 4 N_D \int_{s_0}^{R} ds ~{\rm Im} \Pi(s) (P(s) - 1/s).
\end{equation}

It is convenient to write the contributions to the fit error from the vector
and axial channels separately, as
\begin{equation}
\delta S_{V,A} = 4 N_D \int_{s_0}^{R} ds~ {\rm Im}
\Pi_{V,A}(s) (P(s) - 1/s).
\end{equation}
 In order to bound these contributions, we make the
reasonable assumption that the total spectral weight between $s_0$ and $R$ in
each channel is {\em at most} $50$
percent larger  than the analogous
quantities in ($N_{C}$-rescaled) QCD. That is,
 \begin{eqnarray}
\int_{s_0}^{R} ds~ {\rm Im} \Pi_V(s) < 0.5 N_{TC} m_C^2, \nonumber \\
\int_{s_0}^{R} ds~ {\rm Im} \Pi_A(s) < 0.4 N_{TC} m_C^2.
\end{eqnarray}
{}From Fig. 3, it is clear that
\begin{equation}
|P(s) - 1/s| < 0.01/m_C^2,
\end{equation}
almost everywhere between $s_0$ and $R$. We arrive at the  inequalities,
\begin{eqnarray}
|4 N_D \int_{s_0}^{R} ds~ {\rm Im} \Pi_{V}(s) (P(s) - 1/s)| < 0.02 N_{TC} N_D
\nonumber \\
|4 N_D \int_{s_0}^{R} ds ~{\rm Im} \Pi_{A}(s) (P(s) - 1/s)| < 0.016 N_{TC} N_D.
\end{eqnarray}
 The net maximal fit error is then $0.036 N_{TC} N_D$. It is highly unlikely
 that this bound is actually saturated, requiring as it does a great
conspiracy of the
 vector and axial spectrum to do so. A more realistic estimate would be about
 $0.01 N_{TC} N_D$. Clearly this does not affect any of the  reasoning in the
 body of the paper.

\newpage

\section*{Figure Captions}

\noindent Figure 1: ACD contours in the complex t plane

\noindent Figure 2: Comparison of $P(t)$ and  $1/t$ ($m_C^{-2}$) for MTC
plotted  against $t$ ($m_C^{2}$)

\noindent Figure 3: Comparison of $P(t)$ and  $1/t$ ($m_C^{-2}$) for WTC
plotted  against $t$ ($m_C^{2}$)

\noindent Figure 4: Re$(P(t) - 1/t)$ and Im$(P(t) - 1/t)$ ($m_C^{-2}$),
$t = R e^{i \theta}$, plotted against $\theta$ (radians). $\theta = 0$ is the
physical $t$ axis.

\end{document}